\documentclass[prl,twocolumn,showpacs]{revtex4-1}
\usepackage[latin1]{inputenc}
\usepackage{graphicx}
\usepackage{amssymb}
\usepackage{amsmath}
\usepackage{xspace}
\usepackage{dcolumn}
\usepackage{bm}
\usepackage{color}
\usepackage{float}
\usepackage[extra]{tipa}

\setcitestyle{square,numbers,compress}

\newfont{\tensy}{cmsy10}

\newcommand{\UP}[0]{\uparrow}
\newcommand{\DO}[0]{\downarrow}

\newcommand{\oP}{\hat{P}}
\newcommand{\oQ}{\hat{Q}}

\newcommand{\on}{\hat{n}}

\newcommand{\rmd}{\text{d}}

\newcommand{\bit}{\begin{itemize}}
\newcommand{\eit}{\end{itemize}}

\newcommand{\om}[0]{\omega}

\newcommand{\kF}{k_\text{F}}
\newcommand{\kB}{k_\text{B}}
\newcommand{\nag}{{\phantom{\dag}}}

\newcommand{\rhob}{\overline{\rho}}

\newcommand{\lb}{\lambda_\text{b}}
\newcommand{\ls}{\lambda_\text{s}}

\newcommand{\las}[0]{\langle}
\newcommand{\ras}[0]{\rangle}

\newcommand{\la}[0]{\left\las}
\newcommand{\ra}[0]{\right\ras}
\newcommand{\ket}[1]{\left|#1\ra}
\newcommand{\bra}[1]{\la#1\right|}

\begin{document} 


\title{Interplay of Site and Bond Electron-Phonon Coupling in One Dimension}

\author{Martin Hohenadler}

\affiliation{\mbox{Institut f\"ur Theoretische Physik und Astrophysik,
    Universit\"at W\"urzburg, 97074 W\"urzburg, Germany}}

\begin{abstract}
  The interplay of bond and charge correlations is studied in a
  one-dimensional model with both Holstein and Su-Schrieffer-Heeger (SSH) couplings
  to quantum phonons. The problem is solved exactly by quantum Monte Carlo
  simulations.  If one of the couplings dominates, the ground state is a
  Peierls insulator with long-range bond or charge order. At weak coupling,
  the results suggest a spin-gapped and repulsive metallic phase arising from the
  competing order parameters and lattice fluctuations. Such a phase is absent
  from the pure SSH model even for quantum phonons. At strong coupling,
  evidence for a continuous transition between the two Peierls states is
  presented.
\end{abstract}

\date{\today}

\maketitle

Electron-phonon coupling is at the heart of phenomena such as polaron
formation, the Peierls instability, superconductivity, and relaxation in
nonequilibrium. At the same time, a correct quantum mechanical treatment
poses a serious challenge even for modern numerical methods. The
rich physics arising from the fundamental Holstein density-displacement \cite{Ho59a} or
Su-Schrieffer-Heeger (SSH) hopping-displacement coupling
\cite{PhysRevLett.42.1698} has been studied in great detail \footnote{For a
  recent review of the polaron problem see
  Ref.~\cite{0034-4885-72-6-066501}, and for the half-filled case see
  Refs.~\cite{PhysRevB87.075149,PhysRevB.91.245147}.}. However, their interplay
is so far unexplored \footnote{After submission
of this work, results for the interplay of bond and site coupling in the
polaron problem were published in Ref.~\cite{Berciu2016}.}, even though both
couplings will typically be present in materials \cite{Pouget2016332}.

\begin{figure}[t]
  \includegraphics[width=0.325\textwidth]{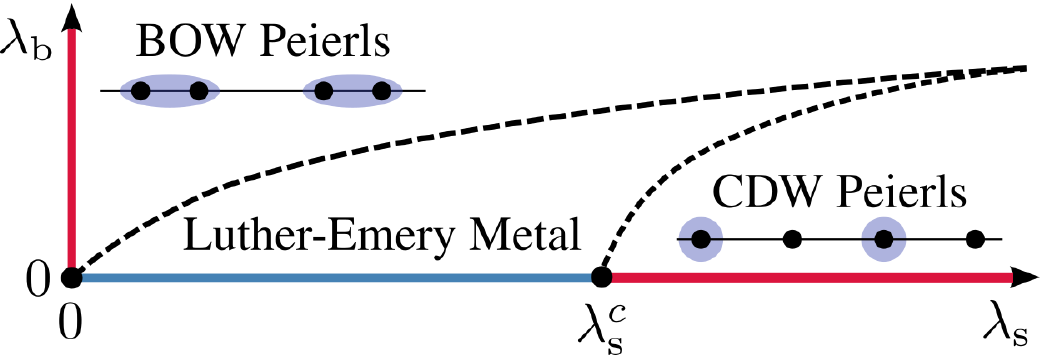}
  \caption{\label{fig:phasesschematic} Schematic phase diagram of the
    model~(\ref{eq:model}). Dots indicate critical points of the Holstein and
    SSH models, dashed lines illustrate the expected phase
    boundaries.}
\end{figure}

For a half-filled band in one dimension with Fermi wave vector $\kF=\pi/2$, a
Holstein coupling causes a Peierls transition to a $2\kF$ charge-density-wave
(CDW) insulator at a critical coupling $\ls^c>0$, as illustrated in
Fig.~\ref{fig:phasesschematic}. In contrast to mean-field theory, quantum
lattice fluctuations destroy the Peierls state for $\ls<\ls^c$ and produce a
spin-gapped metallic phase \cite{JeZhWh99}. On the other hand, the $2\kF$ bond-order wave
(BOW) Peierls state of the (spinful) SSH
model---with alternating weak and strong bonds---is stable for any $\lb>0$
\cite{PhysRevB.27.1680,PhysRevB.67.245103,PhysRevB.73.045106,Barkim2015,PhysRevB.91.245147}.
Hence, in this case, Peierls's theorem \cite{Peierls} holds beyond the adiabatic
limit. Different components of the retarded phonon-mediated interactions give
rise to the spin gap and the Peierls state. Even a qualitatively correct understanding of these models is
beyond the widely used adiabatic and antiadiabatic approximations or even
the bosonization, and instead requires advanced methods
\cite{Barkim2015,PhysRevB.92.245132}.  More realistic models with both
interactions have so far not been studied.

The competition between interactions in the presence of lattice fluctuations
is complex. Whereas metallic behavior arises from competing electron-electron
and Holstein electron-phonon interactions
\cite{ClHa05,0295-5075-84-5-57001}, SSH models only support Peierls and Mott
states \cite{PhysRevB.67.245103,Barkim2015,PhysRevB.91.245147}. In contrast
to the simpler yet intricate extended Hubbard model 
\cite{PhysRevB.45.4027,PhysRevLett.89.236401,Sa.Ba.Ca.04,PhysRevLett.99.216403},
there are no nontrivial integrable limiting
cases in the electron-phonon problem. Two key questions are
if a metallic phase can emerge from the interplay of CDW and BOW order, and
how the transition between the different Peierls states takes place. 

In this Letter, we study this problem by exact quantum Monte Carlo
simulations of the Hamiltonian
\begin{align}\label{eq:model}\nonumber
  \hat{H}
  &=
    -t \sum_{i} \hat{B}_{i}
    +
    \sum_{i,\alpha}
    \left[
    \mbox{$\frac{1}{2M_\alpha}$} \oP^2_{i,\alpha}
    +
    \mbox{$\frac{K_\alpha}{2}$} \oQ_{i,\alpha}^2
    \right]
  \\
  &\phantom{=}
    -
    g_\text{s} 
    \sum_{i} (\on_{i}-1) \hat{Q}_{i,\text{s}} 
    -
    g_\text{b}
    \sum_{i} 
    \hat{B}_{i}
    \oQ_{i,\text{b}}
    \,.
\end{align}
The bond operator
$\hat{B}_{i} = \sum_\sigma(\hat{c}^\dag_{i\sigma} \hat{c}^\nag_{i+1\sigma} +
\hat{c}^\dag_{i+1\sigma} \hat{c}^\nag_{i\sigma})$,
where $\hat{c}^\dag_{i\sigma}$ creates an electron with spin $\sigma$ at
lattice site $i$, and the density operator 
$\on_i = \sum_\sigma c^\dag_{i\sigma} c^\nag_{i\sigma}$. The first term in
Eq.~(\ref{eq:model}) describes the hopping of electrons between neighboring
lattice sites with amplitude $t$. The second term models harmonic
oscillations on site or bond $i$, with $M_{\alpha}$ and $K_{\alpha}$ the mass
and stiffness constant associated with the independent site
($\alpha=\text{s}$) and bond phonons ($\alpha=\text{b}$), respectively. 
The third (fourth) term is a Holstein (SSH) coupling
to site (bond) displacements $\oQ_{i,\text{s}}$
($\oQ_{i,\text{b}}$). Equation~(\ref{eq:model}) reduces to the Holstein model
for $\lb=0$, and to the optical SSH model for $\ls=0$. The latter gives the
same results as the SSH model with acoustic phonons and a coupling
$g_\text{b}\sum_{i} \hat{B}_{i} (\oQ_{i+1,\text{b}} - \oQ_{i,\text{b}})$ but
has no sign problem \cite{PhysRevB.91.245147}.  In terms of the $g_\alpha$,
the coupling constants $\lambda_\alpha$ are given by
$\ls=g_\text{s}^2/(4K_\text{s}t)$ and $\lb=g_\text{b}^2/(K_\text{b}t)$,
respectively. We take the phonon frequencies
$\omega_{0,\alpha}=\sqrt{K_\alpha/M_\alpha}$ to be equal,
$\omega_{0,\alpha}=\omega_0=0.5t$, set $\hbar=\kB=1$, and consider
half-filled ($\las\on_i\ras=1$, chemical potential $\mu=0$) periodic chains
with $L$ sites at inverse temperature $\beta t= t/T = L$.

\begin{figure}[t]
  \includegraphics[width=0.45\textwidth]{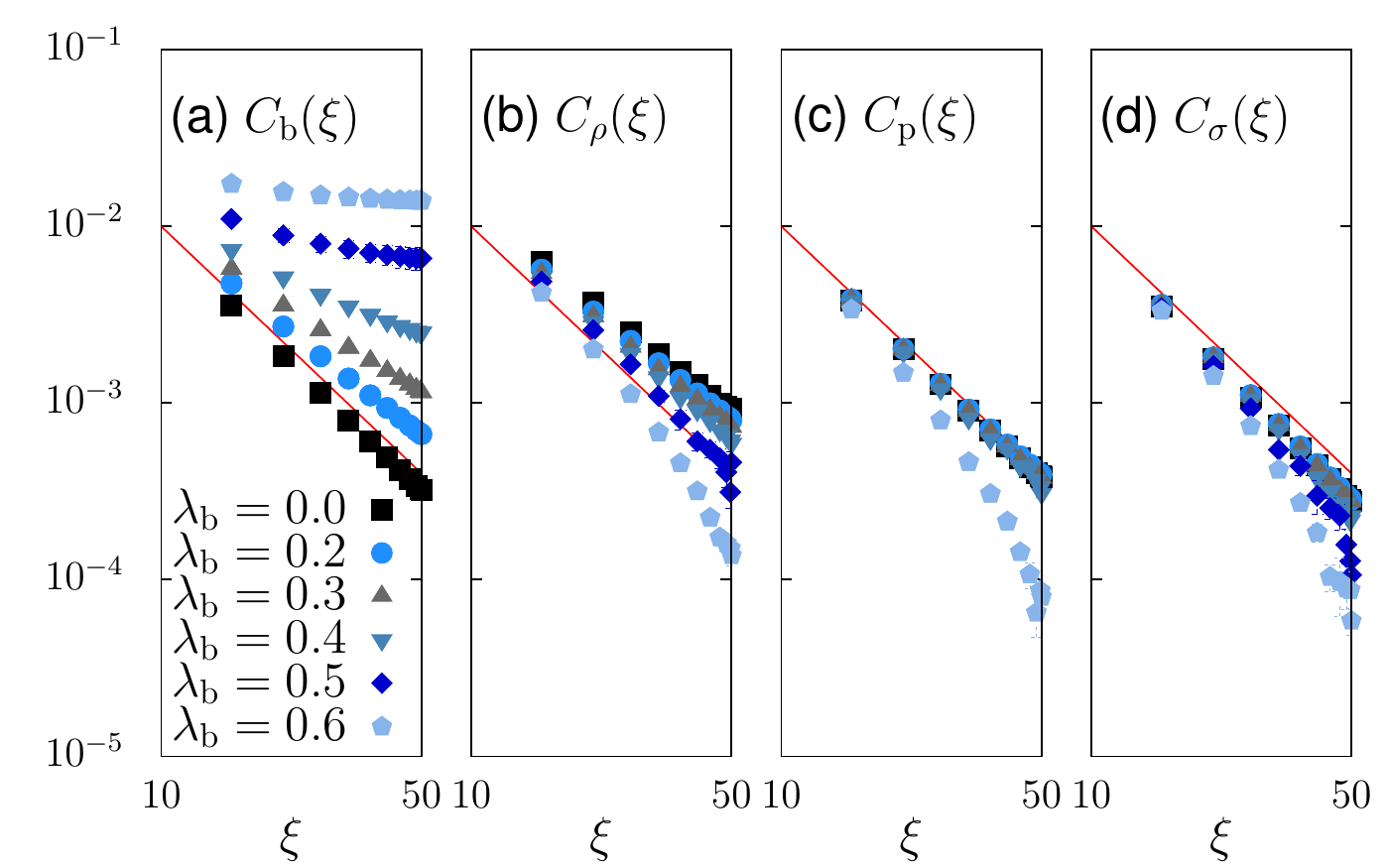}
  \caption{\label{fig:correlatorsweak} (a) Bond, (b) charge, (c) pairing, 
    (d) spin correlations as a function of the conformal distance $\xi$ (see
    text). Here, $\ls=0.15<\ls^c$, $L=50$. Solid lines illustrate a
    $r^{-2}$ decay.}
\end{figure}

Hamiltonian~(\ref{eq:model}) represents a significant challenge due to the
different time scales for the dynamics of electrons and phonons and the
infinite bosonic Hilbert space. It can be simulated with the continuous-time
quantum Monte Carlo method of Ref.~\cite{Rubtsov04} which is based on the
exact summation of a weak-coupling expansion of the partition function. A
key advantage of this method is that its path-integral formulation permits
integrating out the phonons and simulating an effective fermionic model \cite{Assaad07}
that captures the full dynamics of the quantum phonons. The resulting retarded interactions have the form
$S^1_\alpha= -\gamma^2_\alpha \iint_0^\beta \rmd \tau \rmd \tau' \sum_q
\rhob_\alpha(q,\tau) D^0_\alpha(q,\tau-\tau') \rho_\alpha(q,\tau')$,
where $D^0_\alpha$ is the free phonon propagator; $\rho_\alpha(q,\tau)$ is a
Grassmann bilinear corresponding to the Fourier transform of the bond
operator $\hat{B}_i$ (for $\alpha=\text{b}$) or the charge operator
$\hat{n}_i$ (for $\alpha=\text{s}$), respectively. As in previous work, we
used single-vertex updates and Ising spin flips \cite{Assaad07}. To calculate
spectral functions, we applied a maximum-entropy method for the analytic
continuation \cite{Beach04a}.

{\em Theoretical expectations.}---Luttinger liquid theory describes
one-dimensional systems in terms of charge or spin Luttinger parameters
$K_{\rho/\sigma}$ and velocities $v_{\rho/\sigma}$ \cite{Voit94}. Whereas the
metallic phase of the Holstein model is only captured if the momentum and
frequency dependence of the interaction is taken into account
\cite{Barkim2015}, the simple {\it g-ology} representation in terms of
backward ($g_1$), forward ($g_2$), and umklapp ($g_3$) scattering
\cite{Voit94} already reveals important differences to the SSH model. Whereas
$g_2$ terms only renormalize the quadratic theory, $g_1$ and $g_3$ have the
potential to open gaps and/or break symmetries if relevant in the
renormalization group sense \cite{Voit94}. The Holstein coupling gives
$g_\text{1,s}=g_\text{2,s}=g_\text{3,s}=-2g_\text{s}^2/\omega_{0,\text{s}}$,
whereas the SSH coupling leads to
$g_\text{1,b}=-g_\text{3,b}=-2g_\text{b}^2/\omega_{0,\text{b}}$, and
$g_\text{2,b}=0$ \cite{Barkim2015}. In the combined model~(\ref{eq:model})
the backscattering $g_{1}=g_\text{1,s}+g_\text{1,b}$ is always attractive,
whereas the total umklapp matrix element $g_{3}=g_\text{3,s}+g_\text{3,b}$
can be small or zero for suitable $g_\text{s}$ and $g_\text{b}$ (or $\ls$ and
$\lb$).

As in the attractive Hubbard model, any $g_1<0$ opens a spin gap in Holstein
and SSH models \cite{PhysRevB87.075149,PhysRevB.91.245147}.  The relevance of
umklapp scattering for any $\lb>0$ in the SSH model has been attributed to
the vanishing of $g_\text{2,b}$ \cite{Barkim2015}. In contrast, weak umklapp
scattering is irrelevant in the Holstein model and gives a metallic
Luther-Emery phase with a gap for single-particle and spin excitations but
gapless charge excitations
\cite{JeZhWh99,ClHa05,0295-5075-84-5-57001,PhysRevB87.075149,PhysRevB.92.245132}.
The model~(\ref{eq:model}) should permit metallic behavior due to $g_2\neq 0$
and the compensation of $g_\text{3,s}$ and $g_\text{3,b}$.  Similarly,
attractive umklapp scattering is compensated by repulsive contributions from
the electron-electron interaction in the Holstein-Hubbard model
\cite{ClHa05}, whereas all $g_3$ terms are positive in the SSH-Hubbard model
for polyacetylene \cite{Barkim2015}. Finally, the Peierls
states break a discrete Ising symmetry at $T=0$, allowing for long-range order.

{\em Weak site coupling.}---We first consider the impact of the SSH coupling
on the metallic phase of the Holstein model with $\ls=0.15<\ls^c\approx0.25$
\cite{hardikar:245103,0295-5075-84-5-57001,PhysRevB.92.245132}.
Figure~\ref{fig:correlatorsweak} shows the correlators
$C_\text{b}(r) = \las \hat{B}_{r} \hat{B}_0\ras -\las\hat{B}_r\ras^2$,
$C_{\rho}(r) = \las (\on_r - 1) (\on_0-1) \ras$,
$C_\text{p}(r)=\las c^\nag_{r\UP} c^\nag_{r\DO}  c^\dag_{0\UP} c^\dag_{0\DO} \ras$, and
$C_{\sigma}(r) = \las \hat{s}^x_r \hat{s}^x_0 \ras$
($\hat{s}^x_r=c^{\dagger}_{r\uparrow}c^\nag_{r\downarrow}+c^{\dag}_{r\downarrow}
c^\nag_{r\uparrow}$)
as a function of the conformal distance $\xi = L\sin ({\pi r}/{L})$
\cite{Cardybook}. Starting from dominant CDW correlations at $\lb=0$,
CDW (BOW) correlations are suppressed (enhanced) with increasing
$\lb$. At the same time, pairing is slightly enhanced
[Fig.~\ref{fig:correlatorsweak}(c)].  All three channels initially retain a
power-law form, compatible with a metallic phase.  For larger $\lb$ we
observe long-range bond order, and exponentially decaying charge, pairing,
and spin correlations. The different decay of charge (or bond) and spin
correlations implies a spin gap for all $\lb$: In a gapless Luttinger liquid
the $2\kF$ part of all three correlators decays with exponent
$K_\rho+K_\sigma$ \cite{Voit98}. In contrast, in a Luther-Emery liquid, the
exponent is $K_\rho$ for bond/charge ($K_\rho^{-1}$ for pairing) while spin
correlations decay exponentially \footnote{In the bosonization, the dominance
  of bond over charge correlations follows from the different scaling of the
  amplitude under the renormalization group flow
  \cite{PhysRevB.45.4027}.}. The dominance of bond over pairing correlations
indicates effectively repulsive interactions ($K_\rho<1$).

\begin{figure}[t]
  \includegraphics[width=0.45\textwidth]{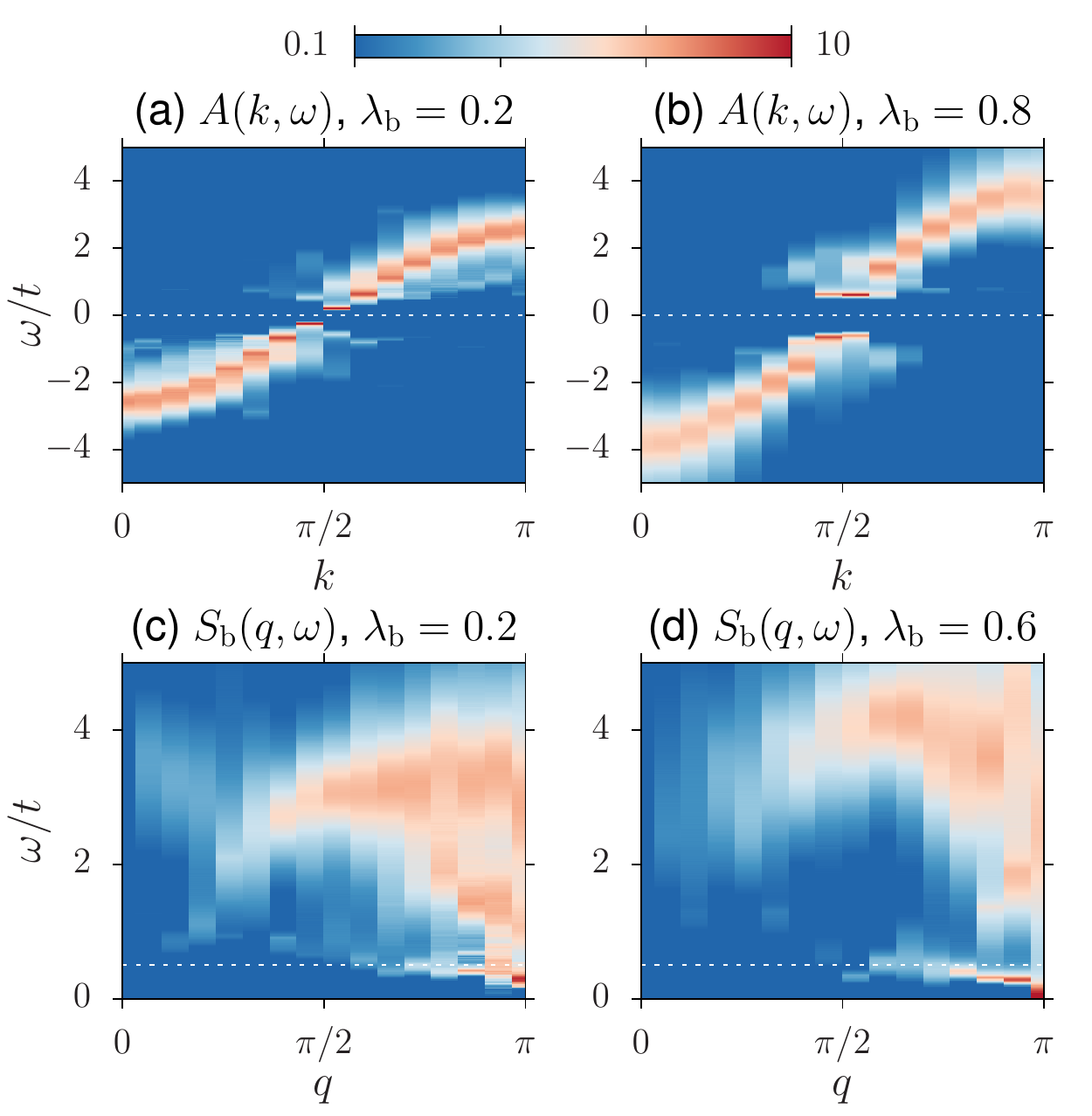}
  \caption{\label{fig:dynamicsweak} (a)--(b) Single-particle spectral
    function, and (c)--(d) dynamic bond structure factor. Here,
    $\ls=0.15<\ls^c$, and $L=30$.  Dashed lines indicate $\omega=0$ and
    $\omega=\omega_0$, respectively.}
\end{figure}

The transition to the BOW-Peierls phase is also reflected in the excitation
spectra shown in Fig.~\ref{fig:dynamicsweak}. For $\lb=0.2$
[Fig.~\ref{fig:dynamicsweak}(a)], the single-particle spectral function
\footnote{{In the Lehmann representation we have
    $A(k,\omega) = \sum_{ij} {|\bra{i} c_{k\sigma} \ket{j}|}^2 (e^{-\beta
      E_i}+e^{-\beta E_j}) \delta(E_j-E_{i}-\om)/Z$,
    where $E_i$ is the energy of the eigenstate $|i\rangle$. $A(k,\omega)$
    was obtained from the Green function $G(\bm{k},\tau)$ by analytic
    continuation.}}  reveals a small spin gap at the Fermi level. The
electron-phonon coupling significantly broadens the excitations outside the
coherent interval $|\omega|<\omega_0$ \cite{Meden94,Hohenadler10a}. A clear
Peierls gap can be seen for $\lb=0.8$ [Fig.~\ref{fig:dynamicsweak}(b)]. The
bond phonon dispersion is visible in the dynamic bond structure factor
$S_\text{b}(q,\om)$
\footnote{{$S_\alpha(q,\om) =\sum_{ij} {|\bra{i} \hat{O}_\alpha(q)
      \ket{j}|}^2 e^{-\beta E_i} \delta(E_{j}-E_i-\om)/Z$,
    where $\hat{O}_\alpha(q)$ is the Fourier transform of either
    $\hat{O}_\text{b}(r)=\hat{B}_r-\las\hat{B}_r\ras$ or
    $\hat{O}_\rho(r)=\on_r-1$.}}, in addition to signatures of the
particle-hole continuum. For $\lb=0.2$ [Fig.~\ref{fig:dynamicsweak}(c)] the
mode is slightly softened near the zone boundary, whereas the complete
softening for $\lb=0.6$ [Fig.~\ref{fig:dynamicsweak}(d)] is consistent with
long-range bond order.

\begin{figure}[b]
  \includegraphics[width=0.45\textwidth]{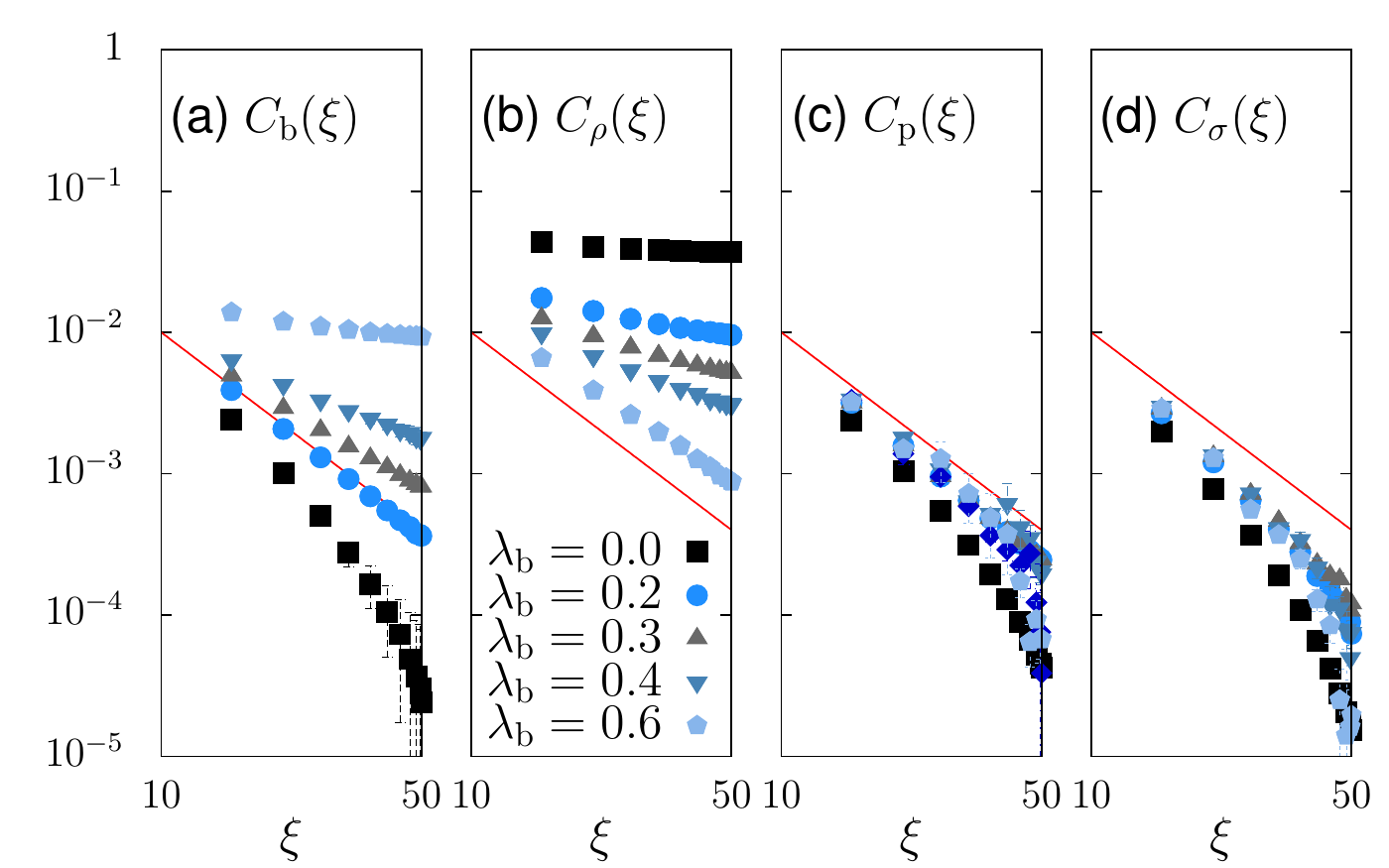}
  \caption{\label{fig:correlatorsstrong} 
    As in Fig.~\ref{fig:correlatorsweak} but for $\ls=0.3>\ls^c$.}
\end{figure}

{\em Strong site coupling.}---We now consider the effect of the bond coupling
on the CDW Peierls state at $\ls=0.3$. For $\lb=0$ we have long-range charge
order [Fig.~\ref{fig:correlatorsstrong}(b)] and exponential bond, pairing,
and spin correlations. A nonzero bond coupling enhances bond correlations,
whereas charge correlations are suppressed until they closely follow the
free-fermion result $1/r^2$ for $\lb=0.5$.  The pairing and spin correlations are enhanced at
intermediate $\lb$, but remain exponential.

\begin{figure}[t]
  \includegraphics[width=0.45\textwidth]{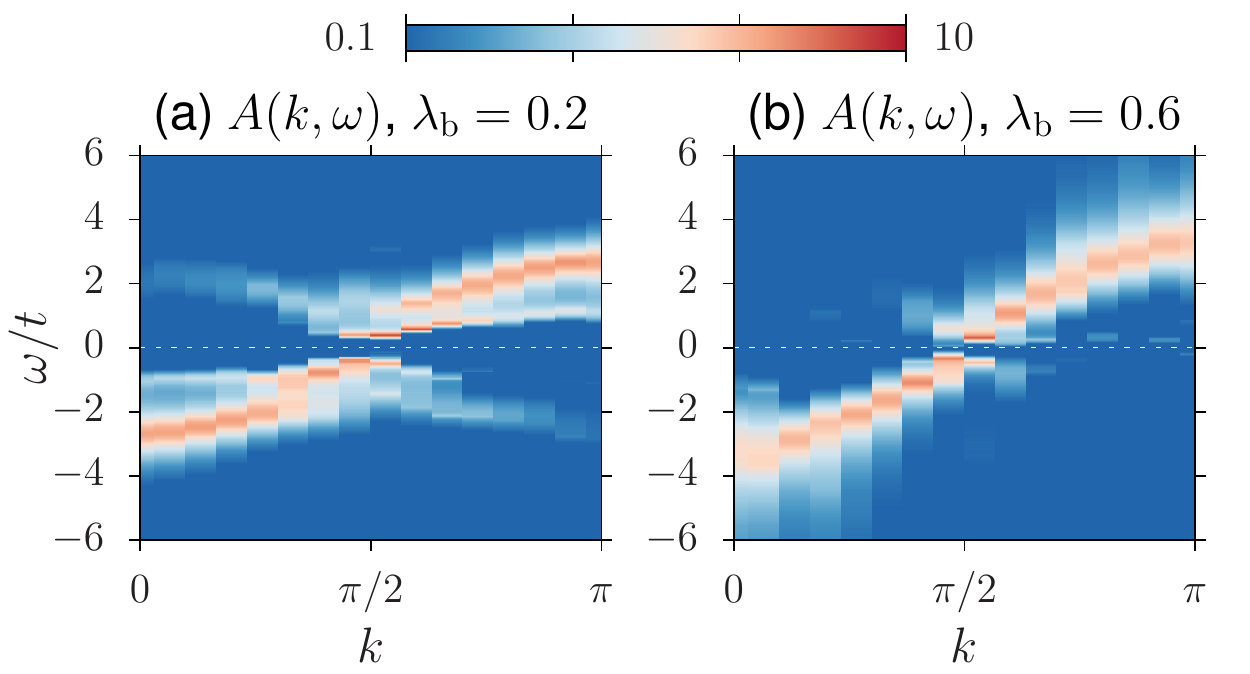}
  \caption{\label{fig:dynamicsstrong} Single-particle spectral function for
    $\ls=0.4$, $L=30$. Here, (a) $\lb=0.2$ and (b) $\lb=0.6$.}
\end{figure}

The single-particle spectral function is shown in
Fig.~\ref{fig:dynamicsstrong} for $\ls=0.4$.  For weak bond coupling
$\lb=0.2$, the spectrum in Fig.~\ref{fig:dynamicsstrong}(a) exhibits the key
features of a Peierls insulator, namely a gap at the Fermi level, backfolded
shadow bands, and soliton excitations inside the mean-field gap
\cite{Vo.Pe.Zw.Be.Ma.Gr.Ho.Gr.00,Hohenadler10a}. Remarkably, for $\lb=0.6$
[Fig.~\ref{fig:dynamicsstrong}(b)], the Peierls features are strongly
suppressed and the spectrum closely resembles that of
Fig.~\ref{fig:dynamicsweak}(a), despite the strong bare couplings. The
corresponding bond
and charge structure factors are discussed below.

\begin{figure}[b]
  \includegraphics[width=0.45\textwidth]{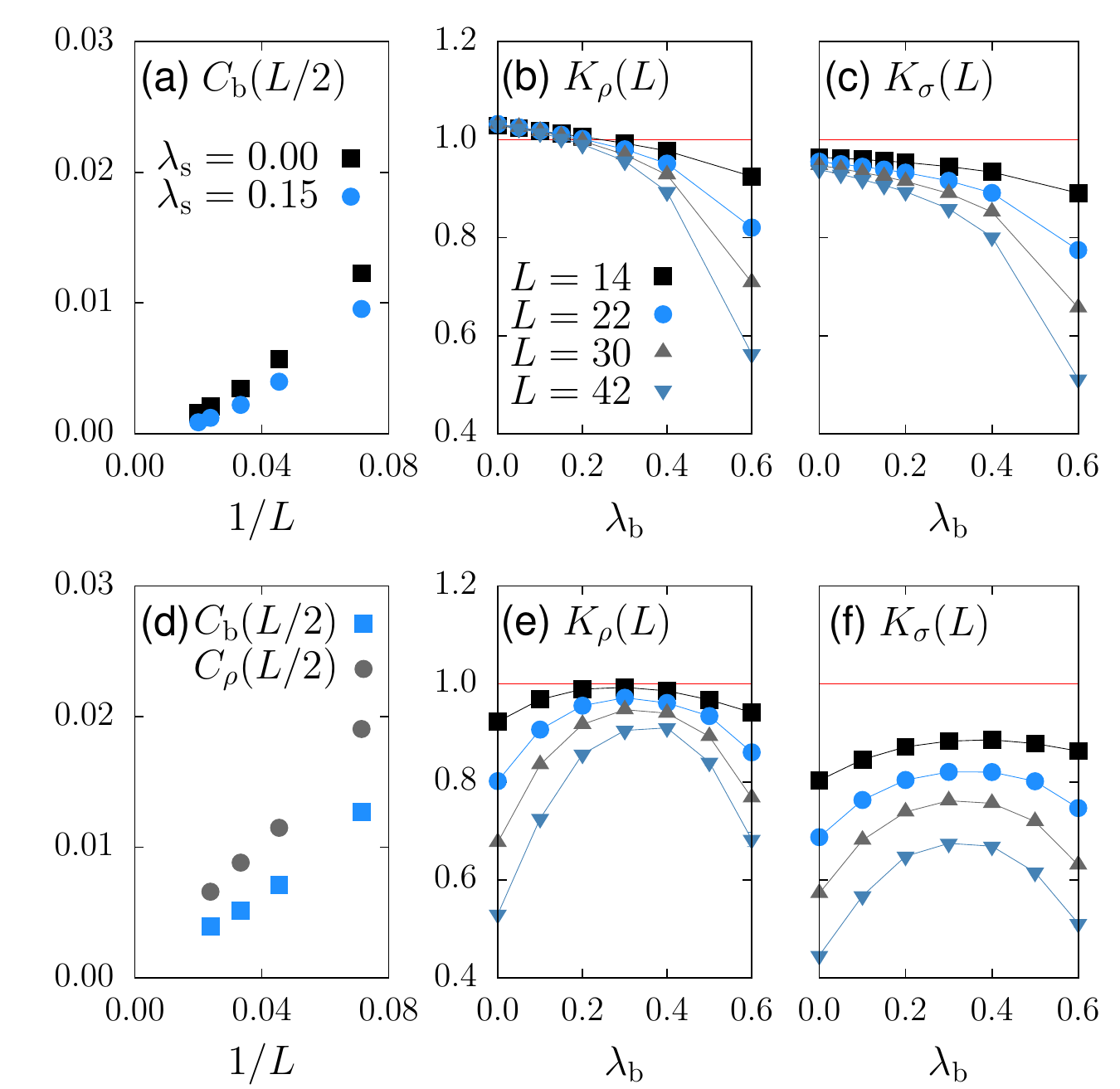}
  \caption{\label{fig:luttinger} (a) Bond correlations for $r=L/2$,
    $\lb=0.1$. (b)--(c), (e)--(f) Finite-size estimates of the Luttinger
    parameters $K_\rho$ and $K_\sigma$. (d) Bond and charge correlations for
    $r=L/2$, $\lb=0.4$.  Here, $\ls=0.15$ in (b)--(c) and $\ls=0.3$ in
    (d)--(f).}
\end{figure}

{\em Metallic phase and CDW-BOW transition.}---To substantiate metallic behavior at weak coupling, we consider the bond
correlations at the largest distance $L/2$, which serve as an order parameter
for the Peierls transition. Figure~\ref{fig:luttinger}(a) illustrates that at
$\lb=0.1$ these correlations are smaller for $\ls>0$ than for $\ls=0$ for
each $L$. While the extrapolation $L\to\infty$ is nontrivial
\cite{PhysRevB.91.245147}, the suppression of $2\kF$ bond order together with
the absence of CDW order in the Holstein model for $\ls=0.15$ and the
power-law decay of bond, charge, and pairing correlations
(Fig.~\ref{fig:correlatorsweak}) all indicate metallic behavior in the
weak-coupling regime of the phase diagram. The Holstein coupling hence
stabilizes the system against the BOW-Peierls transition of the SSH model.

The exact determination of $K_\rho$ and $K_\sigma$ is problematic for
Luther-Emery liquids \cite{PhysRevB.92.245132}. However, even the qualitative
behavior yields important insights. The usual finite-size estimates
$K_{\rho/\sigma}(L)=\pi C_{\rho/\sigma}(q_1) / q_1$ (with $q_1=2\pi/L$) are
shown in Figs.~\ref{fig:luttinger}(b)--(c) for $\ls=0.15$. We have
$K_\rho(L)\gtrsim1$ for small $\lb$ although pairing correlations are
subdominant (Fig.~\ref{fig:correlatorsweak}). This mismatch is due to the
spin gap that causes a logarithmic convergence with $L$
\cite{PhysRevB.92.245132}.  At larger $\lb$, $K_\rho(L)$ is suppressed and
scales to zero for $L\to\infty$. Given the SU(2) spin symmetry of
Hamiltonian~(\ref{eq:model}) we expect $K_\sigma=1$ for a Luttinger liquid
\cite{Voit94}. Instead, $K_\sigma(L)<1$ implies a nonzero
spin gap \cite{PhysRevB.65.155113} even if the latter is too small to be
visible from the spin correlation functions. The results for $K_{\rho/\sigma}(L)$
in Fig.~\ref{fig:luttinger}(b) resemble those for the Holstein model
\cite{ClHa05,PhysRevB.92.245132}, and support a metal-insulator transition at
$\lb^c>0$.

\begin{figure}[t]
  \includegraphics[width=0.45\textwidth]{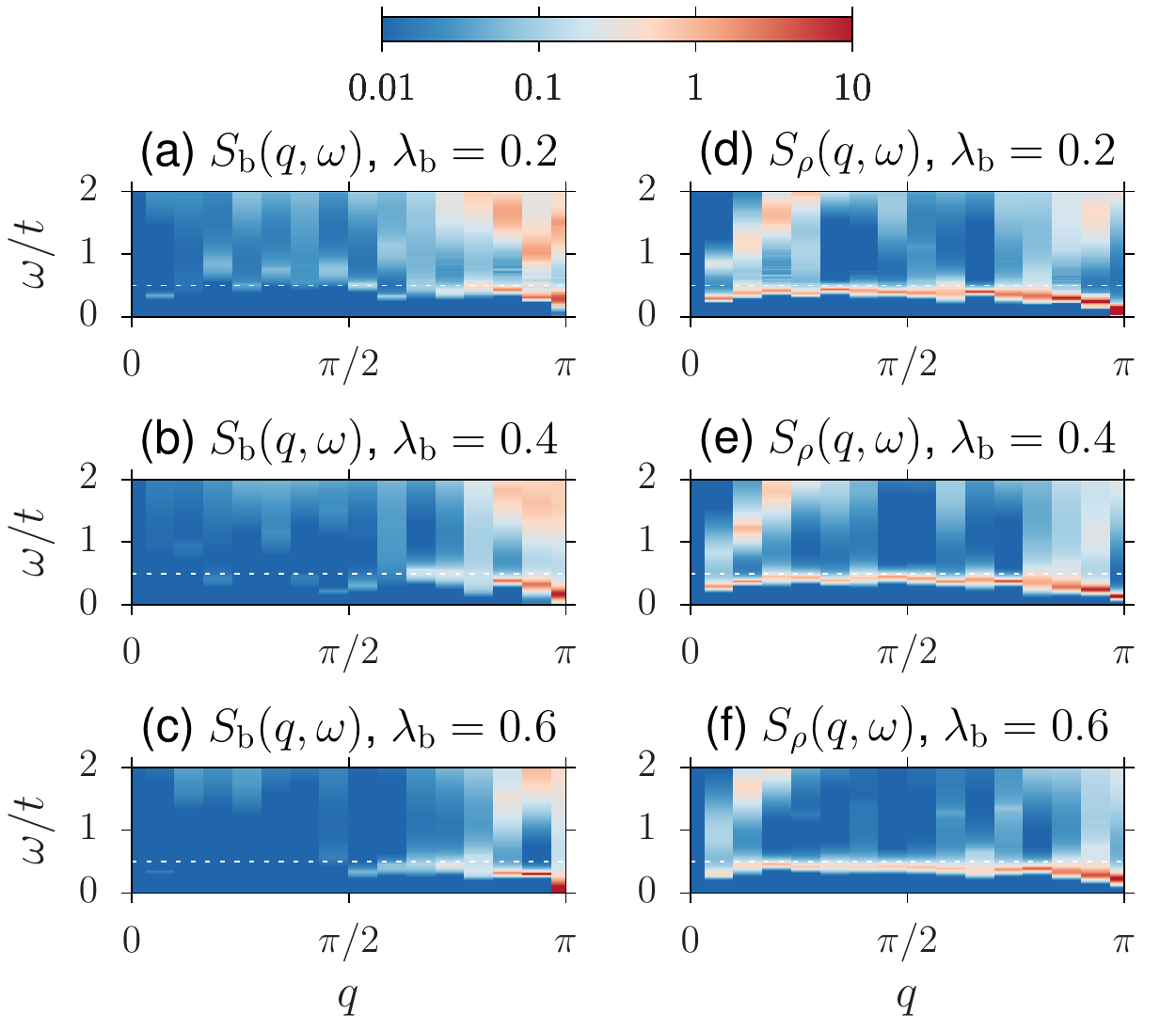}
  \caption{\label{fig:dynamicsphonons} (a)--(c) Dynamic bond structure factor
    and (d)--(f) dynamic charge structure factor for $\ls=0.3$, $L=30$.}
\end{figure}

For a stronger site coupling $\ls=0.3$, the system undergoes a transition
from the CDW Peierls phase to the BOW Peierls phase with increasing $\lb$.
The corresponding suppression (enhancement) of CDW (BOW) correlations was
demonstrated in
Fig.~\ref{fig:correlatorsstrong}. Figure~\ref{fig:luttinger}(d) suggests the
absence of long-range BOW or CDW order at intermediate $\lb=0.4$, and hence
metallic behavior. The value $\lb=0.4$ matches the position of the maximum in
$K_{\rho/\sigma}(L)$ in Figs.~\ref{fig:luttinger}(e)--(f), near which we also
observe the crossover from dominant CDW to dominant BOW
correlations. Furthermore, the single-particle gap is found to take on a
minimum (not shown).  The data for $K_\rho(L)$ in Fig.~\ref{fig:luttinger}(e)
appear to saturate near the maximum---a signature of a continuous phase
transition with a closing of the charge gap \cite{Sa.Ba.Ca.04}. In contrast,
$K_\sigma(L)<1$ and $K_\sigma(L)\to 0$ for $L\to\infty$
[Fig.~\ref{fig:luttinger}(f)] is consistent with a nonzero spin gap across
the transition.  These findings contrast the behavior across the Peierls-Mott
transition in the Holstein-Hubbard model (at which all gaps close and
$K_\rho=K_\sigma=1$ at the critical point \cite{ClHa05,0295-5075-84-5-57001})
and in the SSH-Hubbard model (at which only the spin gap closes
\cite{PhysRevB.67.245103,PhysRevB.91.245147}). The extended Hubbard
model also exhibits CDW and BOW phases, but the intermediate Luther-Emery phase is
restricted to a line \cite{PhysRevB.45.4027,PhysRevLett.89.236401,Sa.Ba.Ca.04,PhysRevLett.99.216403}.

The transition from CDW to BOW order is further reflected in the renormalized
phonon dispersions in Fig.~\ref{fig:dynamicsphonons}.
For weak coupling $\lb=0.2$, the bond phonon mode is slightly renormalized
[Fig.~\ref{fig:dynamicsphonons}(a)] whereas the site phonon mode is soft
[Fig.~\ref{fig:dynamicsphonons}(d)]. At strong coupling $\lb=0.6$, the bond
mode is soft [Fig.~\ref{fig:dynamicsphonons}(c)] while the site mode has
hardened toward $\omega_0$ [Fig.~\ref{fig:dynamicsphonons}(f)]. In between,
Figs.~\ref{fig:dynamicsphonons}(b) and (e), neither mode is soft, consistent
with the absence of long-range order and metallic behavior.

{\em Phase diagram.}---While quantitative phase boundaries are beyond the
scope of the present work, the schematic phase diagram of
Fig.~\ref{fig:phasesschematic} was inferred from the following
observations. Starting from $\ls=0$, the metallic region is expected to grow
because for a larger $\ls$ the umklapp terms cancel at a larger $\lb$. More
physically, stronger CDW correlations require a larger $\lb$ to be
compensated before the BOW Peierls phase can emerge. An increase of the
critical coupling is supported by a maximum in $K_{\rho/\sigma}(L)$ at larger
$\lb^\text{max}\approx 0.55$ for $\ls=0.4$ as compared to $\lb^\text{max}\approx 0.4$ for
$\ls=0.3$.  Similarly, the CDW Peierls phase should appear at larger $\ls$
for $\lb>0$ due to the competing BOW correlations. Finally, the reduced
extent of the metallic phase for $\ls>\ls^c$ was deduced from the observation
of a sharper maximum in $K_\rho(L)$ for the same $L$ at $\ls=0.4$ compared to
$\ls=0.3$. Whether there is a finite metallic phase or a metallic transition
point in this regime has to be answered elsewhere.

{\em Conclusions and outlook.}---We studied a half-filled one-dimensional model of
electrons coupled to site and bond quantum phonons by an exact quantum Monte Carlo method.
The corresponding Holstein and SSH couplings favor different Peierls states.
The results suggest that their competition provides a mechanism to restore
the metallic behavior spuriously absent from the SSH model, in the form of 
a repulsive Luther-Emery phase with a spin gap. The transition between the
two Peierls states appears to be continuous, and the Peierls gaps partially
cancel each other.  A physical picture is that of
either bond or site spin singlets that are ordered in the Peierls phases but
disordered in the metallic phase. While originally motivated by electron-phonon
coupling in materials, the model also represents a generalization of extended
Hubbard models to the case of retarded interactions, and describes the
interplay of spin, charge and lattice fluctuations. 
Directions for future work include the exact
phase diagram with its potential multicritical point, the impact of
Mott physics driven by electron-electron repulsion, as well as 
competing interactions and finite-temperature phase transitions in 
two dimensions.

{\begin{acknowledgments}%
    The author gratefully acknowledges the Gauss Centre for Supercomputing
    e.~V. (www.gauss-centre.eu) for funding this project by providing
    computing time on the GCS Supercomputer SuperMUC at Leibniz
    Supercomputing Centre (LRZ, www.lrz.de).  Discussions with F.~F. Assaad are
    acknowledged. This work was supported by the DFG through SFB 1170 ToCoTronics.
  \end{acknowledgments}}


%

\end{document}